LAPP-EXP 2004-11
December 2004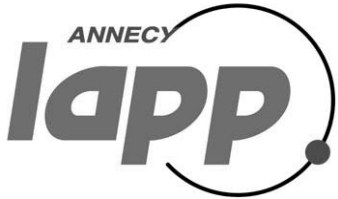

# The OPERA experiment

# H. Pessard

LAPP-IN2P3-CNRS
9 chemin de Bellevue - BP. 110
F-74941 Annecy-le-Vieux CedexFor the OPERA Collaboration

**Invited talk at the 32nd International Conference on High Energy Physics ICHEP'04,
Beijing, China, 16-22 August 2004**

# THE OPERA EXPERIMENT

HENRI PESSARD[*]

*LAPP-IN2P3-CNRS, 9 Chemin de Bellevue, BP 101, 74941 Annecy-le-Vieux Cedex, France*

OPERA is a neutrino oscillation experiment designed to perform a $\nu_\tau$ appearance search at long distance in the future CNGS beam from CERN to Gran Sasso. It is based on the nuclear emulsion technique to distinguish among the neutrino interaction products the track of a $\tau$ produced by a $\nu_\tau$ and its decay tracks. The OPERA detector is presently under construction in the Gran Sasso underground laboratory, 730 km from CERN, and will receive its first neutrinos in 2006. The experimental technique is reviewed and the development of the project described. Foreseen performances in measuring $\nu_\tau$ appearance and also in searching for $\nu_e$ appearance are discussed.

## 1. The CNGS program and OPERA

### 1.1. *Long baseline projects in Europe*

Following the discovery of neutrino oscillations by Super-Kamiokande in 1998 with atmospheric neutrinos, projects using accelerator neutrinos have developed in Japan and the USA to measure the disappearance of $\nu_\mu$ at long distance. In Europe, long baseline projects focused on the appearance of $\nu_\tau$ in a $\nu_\mu$ beam have led to the construction of the CNGS (CERN Neutrinos to Gran Sasso) beam at CERN. Their physics objective is to prove explicitly the $\nu_\mu$–$\nu_\tau$ nature of the atmospheric oscillation and check the $\Delta m^2$ value in this channel. Searching for $\nu_\mu$–$\nu_e$ oscillations in this beam will also provide a window of opportunity to measure $\theta_{13}$ before the next generation of dedicated experiments. Two detectors, OPERA and ICARUS, will be placed in the Gran Sasso underground laboratory, 730 km from CERN.

The CNGS beam is a wide band neutrino beam optimized for $\tau$ appearance with a mean neutrino energy of 17 GeV. SPS beam extraction, target station with focusing horns, decay tunnel 1 km long, are new. The civil engineering is finished, hadron stop and decay tube have been successfully installed. The beam construction is well on schedule and first beams to Gran Sasso are expected in May 2006. With a foreseen intensity of $4.5 \; 10^{19}$ protons of 400 GeV on target per year (assuming 200 days of operation and 55% overall efficiency), the number of neutrino interactions expected in the 1.8 kton OPERA target is about 6200/year and the number of $\nu_\tau$ CC interactions 27/year (for $\Delta m^2 = 2.4 \; 10^{-3} \; \text{eV}^2$). An intensity upgrade of a factor 1.5, presently under study, seems feasible at a limited cost [1].

### 1.2. *The OPERA detector principles*

The OPERA experiment is based on the direct observation of the $\tau$ decay topology [2], by means of nuclear photographic emulsions. The detection of the $\tau$ decay kink in $\nu_\tau$ CC events has been studied for the 1-prong $\tau$ decay modes (85% total branching ratio) associating emulsion films for tracking with high Z plates as targets. This technique was used in the $\nu_\tau$ discovery by DONUT in 2000 [3]. In OPERA, emulsion films made of a plastic base with two emulsion layers of 45μm are alternated with lead plates 1mm thick. The necessity to keep film alignment within a micron while achieving a target mass of 1800 tons leads to a highly modular target. The basic target unit is a brick of 10.2 x 12.7 x 7.5 cm made of 56 lead plates and 57 emulsion films. There are about 200 000 bricks in total.

---

[*] On behalf of the OPERA Collaboration.

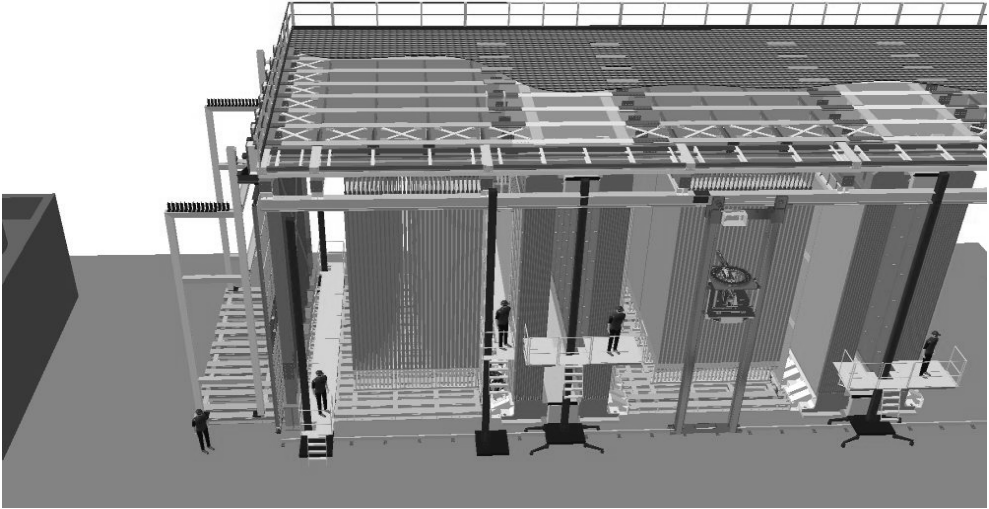

Figure 1. Sketch of the OPERA detector showing the spectrometer magnets and the structure supporting the two target blocks. On the right, a brick manipulator is represented working on the side of the second target block.

After emulsion film scanning performed by automated microscopes, charged tracks will be reconstructed allowing for vertex and decay kink finding. In addition, for low energy tracks, momentum measurement using MCS in the lead plates and $\pi/\mu$ separation using dE/dx are possible. Electromagnetic showers developing in the 10 $X_0$ bricks can be measured, electrons identified and separated from $\gamma$s and $\pi^0$s.

Interspersed between the target bricks stacked in vertical walls, electronic detectors allow to localize on line neutrino interactions and select the corresponding bricks. The Target Tracker (TT) planes consist of horizontal and vertical scintillator strips read out by 64-channel PMTs via WLS fibers, their DAQ electronic cards define the trigger.

There are two target blocks (Figure 1) of 31 brick planes and 31 TT planes. To identify muons and measure their momentum and charge to fight charm background, each target block is followed by a spectrometer composed of a dipolar magnet filled with 22 RPC planes and complemented by 6 sections of drift tubes for precision tracking through the magnetic field.

During data taking, after brick selection and muon measurement by the electronic detectors when a neutrino interaction is tagged, bricks will be extracted daily by two automated manipulators. After brick cosmic ray exposure near the surface to produce alignment tracks and film development, scanning stations in Japan and in Europe will operate quasi on-line to complete fully data acquisition with vertex location and $\tau$ decay kink detection. For $\tau$ decay candidate events, further scanning measurements will be made to improve particle identification and momentum assignment.

In addition to 'long decays' where the $\tau$ track is seen in an emulsion film and kink angles larger than 20mrad (~3mrad error) are selected, 'short decays' where the $\tau$ decays in the lead plate he was produced are also considered for DIS events. Then impact parameters of 5 to 20μm are selected depending on vertex position in the lead (0.3 to 0.6 μm error).

## 2. Status of OPERA construction

### 2.1. *OPERA spectrometers*

The installation of OPERA in Hall C of the Gran Sasso underground laboratory started in May 2003 with the construction of the spectrometers. The magnets, made of 5 cm thick iron slabs, are 10 m high and 9 m wide and weight about 1000 tons. The first magnet was completed in June 2004, filled with tested RPCs. With the Precision Trackers (PT) added, the expected Δp/p is better than 25%, the μ-ID efficiency larger than 95% and the charge confusion less than 0.3%. Construction of the second magnet has started and its completion is foreseen for April 2005.

RPCs are Bakelite chambers, 462 per magnet (1540 m$^2$). Demanding RPC quality tests led to an acceptance level of 70%. PT planes are composed of four layers of staggered drift tubes, 8m long and 38 mm diameter. Full height prototype measurements show an efficiency better than 99% and a resolution better than 0.3 mm when 0.5 mm is required. PT installation will start in spring 2005.

## 2.2. Target sections

In September 2004, the assembly of the first target trackers is starting. The target installation, totaling 6000 m$^2$ of plastic scintillator strips, each plane alternating with a plane of brick wall support, will last until December 2005. Electronics will be cabled and commissioned during the same period.

The bricks themselves will be assembled by an automated machine at a rate of two bricks per minute and 4800 bricks per week during one year, starting in September 2006. A brick manipulator system will load the detector at the same rate so that the first target section will be filled in March and the second in September 2006. Manipulators are in construction while the brick machine is being prepared by an industrial company. Among brick components, the lead plates have to be produced with a precision of 10μm in thickness and will come from an industrial manufacturer. The emulsion films are developed by Fuji jointly with Nagoya University. Of the 13 millions films, 20% are produced. Film refreshing procedures to erase cosmic ray tracks are being applied in Tono mine (Japan) before delivery at Gran Sasso.

## 2.3. Emulsion scanning preparation

R&D to speed up automatic scanning of emulsions to 20 cm$^2$ per hour or more is actively pursued both in Japan and in Europe. The Japanese approach is based on dedicated hardware and hard coded algorithms, using continuous movement of the film stage and very fast CCD cameras (3000 frames per second). The European labs rely on software algorithms and on commercial hardware developing with the video PC market. They recently reach the goal of 20 cm$^2$/hr. Japanese systems are expected to attain soon similar performances.

## 3. OPERA physics performances

Efficiencies have been calculated for the various categories of τ-decays. They include branching ratios together with losses in brick finding, kink finding, event selection, μ identification and connection. The total efficiency amounts to 9.1%, in considerable improvement from the time where only long decays of τ to μ and electron were considered. This figure was then 5.6%. Other efficiency improvements are being studied. The inclusion of the τ → 3h decay mode could provide an additional 1.0% to the total efficiency. Another study shows that the brick finding efficiency (BFE) could be increased by 10% with improved analysis and brick extraction strategy.

The main background sources are charmed decays, hadron re-interactions and large angle μ scattering. Possible improved π/μ separation using dE/dx in emulsions would reduce the charm background by 40%. Including a re-evaluation of charm production using new CHORUS data., the charm background could go down to 0.28 event. The large angle μ scattering background is an upper limit from past measurements and could be 5 times lower according to calculations including nuclear form factors. This reduction will be checked in a forthcoming test-beam. Table 1 details the backgrounds and Table 2 compares the total background to the expected signal.

Table 1. Expected number of background events in OPERA in 5 years with nominal beam. Numbers in parentheses correspond to possible background reductions.

|  | $\tau \to e$ | $\tau \to \mu$ | $\tau \to h$ | Total |
|---|---|---|---|---|
| Charm decay | .21 | .01 | .16 | .38 (.28) |
| Large angle μ |  | .12 |  | .12 (.02) |
| Hadron re-int. |  | .09 | .12 | .21 (.21) |
| Total / channel | .21 | .22 | .28 | .71 (.52) |

Table 2. Expected number of signal and background events in normal, improved efficiency, and reduced background conditions.

|  | Signal | | | BG |
|---|---|---|---|---|
| $\Delta m^2$ ($10^{-3}$ eV$^2$) | 1.9 | 2.4 | 3.0 |  |
| OPERA nominal | 6.6 | 10.5 | 16.4 | 0.7 |
| w. improved eff. | 8.0 | 12.8 | 19.9 | 1.0 |
| w. BG reduction |  |  |  | 0.8 |

Figure 2 presents the 4σ discovery probability and the 90% CL sensitivity limit of OPERA for the possible improvements discussed above and shows the importance of the expected beam intensity upgrade.

OPERA sensitivity has also been estimated in a search for $\nu_\mu$–$\nu_e$ appearance in the CNGS, looking for an excess of $\nu_e$ CC events [4]. The dominant background from the contamination in $\nu_e$ of the beam (0.8%) and contributions from $\nu_\mu$–$\nu_\tau$ with $\tau \to e$ and from $\nu_\mu$ NC and CC interactions can be reduced using visible energy, missing transverse energy and electron energy.

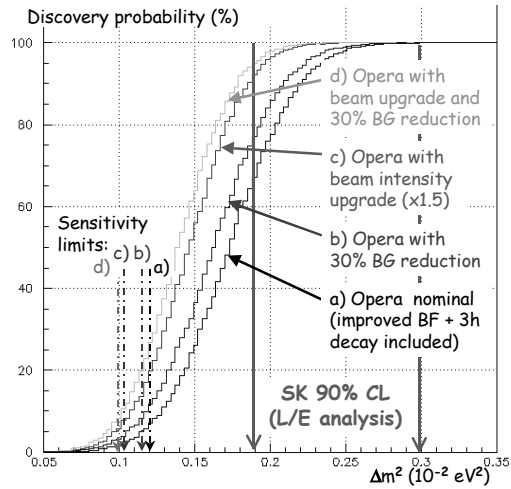

Figure 2. OPERA sensitivity limit and 4σ discovery probability as a function of $\Delta m^2$ for different conditions.

Table 3. Expected number of signal and background events in a search for in 5 years.

| $\theta_{13}$ | $\nu_e$ CC beam | $\tau \to e$ $\nu_\mu$–$\nu_\tau$ | $\nu_\mu$ CC | $\nu_\mu$ NC | $\nu_\mu$–$\nu_e$ signal |
|---|---|---|---|---|---|
| 9° | 18 | 4.5 | 1.0 | 5.2 | 9.3 |
| 7° | 18 | 4.6 | 1.0 | 5.2 | 5.8 |
| 5° | 18 | 4.6 | 1.0 | 5.2 | 3.0 |

Table 3 shows the expected signal and backgrounds after 5 years with nominal beam. The limit obtained at 90% confidence level on $\theta_{13}$ is 7.1°, in significant improvement relative to the present CHOOZ limit [5]. It would reduce to 6.4° with the CNGS beam intensity increase.

## 4. Conclusion

The CNGS beam is well on schedule. OPERA construction is on-going and should be ready to take data by 2006. Great progress has been accomplished towards a successful realization of the experiment. Work is continuing to improve the OPERA sensitivity by reducing the background and increasing the efficiency. The beam intensity upgrade will have the similar effect of securing the $\nu_\tau$ appearance observation even for very low values of $\Delta m^2$.

## References

1. R. Cappi et al., *CERN-SL-2001-032*.
2. OPERA Collaboration, M. Guler et al., Experiment proposal, *CERN-SPSC-2000-028*; Status report on the OPERA experiment, *CERN-SPSC-2001-025*.
3. K. Kodama et al., *Phys. Lett.* **B504**, 218 (2001).
4. M. Komatsu, P. Migliozzi and F. Terranova, *J. Phys.* **G29**, 443 (2003)
5. CHOOZ Collaboration, M. Apollonio et al., *Phys. Lett.* **B466**, 415 (1999).